\documentclass[floats,floatfix,showpacs,amssymb,prd,twocolumn,superscriptaddress,nofootinbib]{revtex4-1}

\usepackage{amssymb,amsmath,verbatim,mathtools,needspace,enumitem,etoolbox,graphicx,physics,microtype,afterpage,bm}

\usepackage[dvipsnames, usenames]{xcolor}
\usepackage[normalem]{ulem}
\usepackage{soul}

\definecolor{linkcolor}{rgb}{0.0,0.3,0.5}
\usepackage[unicode, colorlinks=true, linkcolor=linkcolor, citecolor=linkcolor, filecolor=linkcolor,urlcolor=linkcolor, pdfusetitle]{hyperref}
\usepackage[all]{hypcap}
\usepackage[T1]{fontenc}
\usepackage[utf8]{inputenc}
\usepackage{tabularx}
\usepackage{float}
\allowdisplaybreaks
\usepackage{tensor}

\def\a{\alpha}
\def\b{\beta}

\def\d{\delta}

\def\g{\gamma}

\def\m{\mu}
\def\n{\nu}

\def\r{\rho}
\def\s{\sigma}

\def\tns{\tensor}

\newcommand{\be}{\begin{equation}} 
\newcommand{\ee}{\end{equation}}
\newcommand{\beq}{\begin{equation}} 
\newcommand{\eeq}{\end{equation}}
\newcommand{\bea}{\begin{equation}\begin{aligned}} 
\newcommand{\eea}{\end{aligned}\end{equation}}
\newcommand{\ba}{\begin{eqnarray}}
\newcommand{\ea}{\end{eqnarray}}

\usepackage{orcidlink}

\usepackage{csquotes}
\definecolor{tclr}{RGB}{103,103,246}

\begin{document}

\title{Bimetric Starobinsky model}

\author{Ioannis~D.~Gialamas\orcidlink{0000-0002-2957-52765}}
\email{ioannis.gialamas@kbfi.ee}
\affiliation{Laboratory of High Energy and Computational Physics, 
National Institute of Chemical Physics and Biophysics, R{\"a}vala pst.~10, Tallinn, 10143, Estonia}

\author{Kyriakos~Tamvakis\orcidlink{0009-0007-7953-9816}}
\email{tamvakis@uoi.gr}
\affiliation{Physics Department, University of Ioannina, 45110, Ioannina, Greece}


\begin{abstract} \noindent
The bimetric theory of gravity is an extension of general relativity that describes a massive spin-$2$ particle in addition to the standard massless graviton. The theory is based on two dynamical metric tensors with their interactions constrained by requiring the absence of the so-called {\textit{Boulware-Deser ghost}}. It has been realized that the quantum interactions of matter fields with gravity are bound to generate modifications to the standard Einstein-Hilbert action such as quadratic curvature terms. Such a quadratic Ricci scalar term is present in the so-called Starobinsky model which has been proven to be rather robust in its inflationary predictions. In the present article we study a generalization of the Starobinsky model within the bimetric theory and find that its inflationary behavior stays intact while keeping all consistency requirements of the bimetric framework. The interpretation of the massive spin-2 particle as dark matter remains a viable scenario, as in standard bigravity.
\end{abstract}

\maketitle

\section{Introduction}

 A common working assumption of modern considerations of gravity is that the quantum aspects of gravitation can be ignored for energies below the Planck energy of $10^{19}\, {\rm GeV}$ and, therefore, gravity can be treated classically, in contrast to the full quantum character of Standard Model matter interactions. However gravitating matter fields through their quantum interactions are expected to introduce modifications to the standard general relativity (GR) action with cosmological implications. Such are higher power terms of the curvature as in the case of the Starobinsky model~\cite{Starobinsky1980}, which features a quadratic term of the Ricci scalar curvature. An interesting aspect of the Starobinsky model is the existence of an additional scalar gravitational degree of freedom that plays the role of the inflaton leading to quite robust inflationary predictions in agreement with present observational data~\cite{Planck:2018jri, BICEP:2021xfz}.

 Bimetric gravity (or bigravity)~\cite{Hassan:2011zd}, an extension of GR positing two dynamical metric tensors, each appearing in the action with its own scalar curvature term and interacting through a potential tailored to guarantee consistency, came out of efforts to formulate a theory of massive spin-$2$ particle. These efforts have a rather long history, starting with the linear theory of Fierz and Pauli~\cite{Fierz:1939ix}, which, after circumventing the so-called van Dam-Veltman-Zakharov discontinuity~\cite{vanDam:1970vg, Zakharov:1970cc} by its nonlinear generalization by Vainshtein~\cite{Vainshtein:1972sx}, came to face the existence of ghosts as shown by Boulware and Deser~\cite{Boulware:1972yco}. The important feature of bimetric gravity is that it has been shown to be free of Boulware-Deser ghosts.

 In the present article we consider the standard bimetric action and add to it quadratic Ricci curvature terms $\sqrt{-g}R^2(g)$ and $\sqrt{-f}R^2(f)$ associated with each of the two independent metric tensors $g_{ \mu\nu}$ and $f_{ \mu\nu}$, expected to be generated by quantum interactions of matter fields with each metric and scaled by the corresponding gravitational scales. The action is transformed to the Einstein frame via a Weyl rescaling of each metric where the scalar modes generated by the presence of the quadratic terms acquire explicit kinetic terms. Next we consider the linearized  limit of the theory where the two metrics correspond to perturbations around a common background and analyze the necessary constraints on the energy-momentum tensor of each scalar field. We find that the set of Einstein's equations are satisfied imposing a constraint on the parameters that corresponds to the scalar energy-momentum tensors being proportional. The linear theory describes the standard massless graviton, a Fierz-Pauli spin-$2$ field of mass $m_{\rm FP}$ and  a pair of scalars that leads dynamically to an inflaton potential identical to the standard Starobinsky inflationary potential. It is interesting that all constraints can be met with the so-called {\textit{minimal choice}} of $\beta_n$ parameters of the standard bimetric theory~\cite{Hassan:2011vm}. Nevertheless, more general three-parametric choices are possible allowing for $m_{\rm FP}$ values much below the gravitational scale, which could, in principle, be more interesting phenomenologically, as the extra spin-2 particle that results from the bimetric action is a possible dark matter candidate~\cite{Aoki:2014cla, Aoki:2016zgp, Babichev:2016hir, Babichev:2016bxi, Marzola:2017lbt, Manita:2022tkl, Kolb:2023dzp}. The implications of our findings are noteworthy, as they suggest that the extended bimetric gravity theory, augmented with quadratic Ricci scalar terms, holds considerable promise in effectively describing not only dark matter but also cosmic inflation. By encompassing both of these fundamental aspects within a single theoretical framework, this theory opens up new avenues for investigating the intertwined nature of gravity, dark matter, and the early Universe.
 
 The paper is organized as follows. In Sec.~\ref{the_model} we set up the model and derive the Einstein-frame action and the resulting equations of motion. In Sec.~\ref{linearization} we consider the linearized limit of the model and analyze the arising constraints. In Sec.~\ref{the_scalar_system} we analyze the dynamics of the system of scalars, and in Secs.~\ref{infl_potential} and~\ref{inflation} we derive the associated inflationary behaviour. Finally, in Sec.~\ref{conclusions} we state briefly our conclusions.

\section{The model}
\label{the_model}
We start by extending the standard  bimetric action to include quadratic terms of the Ricci curvature scalars corresponding to each metric, namely,
\begin{align}
\label{eq:act_1}
\mathcal{S}=&\int{\rm d}^4x\Big\{ m_g^2\sqrt{-g}\left(R(g)+\frac{\gamma}{2}R^2(g)\right)\,+\,\nonumber
\\&\,m_f^2\sqrt{-f}\left(R(f)+\frac{\delta}{2}R^2(f)\right)+ 2m_g^2{m^2}\sqrt{-g}V(\sqrt{\Delta})\Big\}\,,
\end{align}
where $g_{ \mu\nu}$ and $f_{ \mu\nu}$ are the two metric tensors and $\tns{\Delta}{^\m_\n}=g^{\m\r}f_{ \r\n}$. The interactions are included in the potential $V(\sqrt{\Delta})$ and its special form is dictated by the consistency of the theory to be
\be 
\label{eq:potV}
V\left(\sqrt{\Delta}\right)\,=\,\sum_{n=0}^4\beta_n\,e_n\left(\sqrt{\Delta}\right)\,,\ee
where $\beta_n$ are free parameters and $e_n\left(\sqrt{\Delta}\right)$ are the elementary symmetric polynomials of
the square-root matrix $\sqrt{\Delta}$, which are given by
\be 
e_n\left(\sqrt{\Delta}\right)=\frac{(-1)^{n+1}}{n}\sum_{k=0}^{n-1}(-1)^k {\rm Tr}\left(\sqrt{\Delta}^{n-k}\right)e_k\left(\sqrt{\Delta}\right)\,, \label{eq:polyn}
\ee 
starting with $e_0\left(\sqrt{\Delta}\right)=1$. The scale $m$ appearing in the action~\eqref{eq:act_1} is a redundant scale expressing the overall scale of the parameters $\beta_n$, usually taken to be $m=m_f$. An equivalent form of the action (\ref{eq:act_1}) can be written in terms of two auxiliary scalars $\Omega^2$ and $\overline{\Omega}^2$ as
\begin{align}
\mathcal{S} = \int {\rm d}^4x\bigg\{ & m_g^2\sqrt{-g}\Omega^2R(g)+m_f^2\sqrt{-f}\overline{\Omega}^2R(f) \nonumber
\\ &  -\frac{m_g^2}{2\gamma}\sqrt{-g}\left(\Omega^2-1\right)^2-\frac{m_f^2}{2\delta}\sqrt{-f}\left(\overline{\Omega}^2-1\right)^2 \nonumber
\\ & +2m^2m_g^2\sqrt{-g}\sum_{n=0}^{4}\beta_ne_n(\sqrt{\Delta})\bigg\}\,.
\label{act_2}
\end{align}
Performing a double Weyl rescaling of the metric tensors
\be g_{ \mu\nu}\rightarrow \Omega^{-2}g_{ \mu\nu},\,\,\,\,\,\,f_{ \mu\nu}\rightarrow\overline{\Omega}^{-2}f_{ \mu\nu}\,,\,\,
\ee
we transform the action to the Einstein frame and obtain
\begin{align}
\mathcal{S} &=  \int {\rm d}^4x \bigg\{  m_g^2\sqrt{-g} R(g)+m_f^2\sqrt{-f} R(f)  \nonumber
\\ & -\frac{1}{2}\sqrt{-g}(\nabla\phi)^2-\frac{1}{2}\sqrt{-f}(\nabla\overline{\phi})^2 \nonumber
\\ &  -\frac{m_g^2}{2\gamma}\sqrt{-g}\left(1-e^{-\frac{\phi}{m_g\sqrt{3}}}\right)^2-\frac{m_f^2}{2\delta}\sqrt{-f}\left(1-e^{-\frac{\overline{\phi}}{m_f\sqrt{3}}}\right)^2 \nonumber
\\ &  +2m^2m_g^2\sqrt{-g}\,\sum_{n=0}^{4}\beta_ne^{\frac{(n-4)\phi}{2m_g\sqrt{3}}}e^{-\frac{n\overline{\phi}}{2m_f\sqrt{3}}}e_n(\sqrt{\Delta})\bigg\}\,,
\label{act_3}
\end{align}
where $\Omega^2=e^{\frac{\phi}{m_g\sqrt{3}}} \quad \text{and} \quad\overline{\Omega}^2=e^{\frac{\overline{\phi}}{m_f\sqrt{3}}}$.
 It is evident that each $R^2$ term\footnote{See~\cite{Nojiri:2012zu, Nojiri:2012re, Nojiri:2015qyc} for cosmological applications in the context of $F(R)$ bimetric gravity.} introduces an extra propagating scalar degree of freedom. However, it is important to note that in the metric-affine formulation of gravity, where the metric tensor and the connection are treated as independent variables, no additional dynamical degree of freedom emerges. The influence of an $R^2$ term in the bimetric theory within the framework of metric-affine gravity has been investigated in~\cite{Gialamas:2023aim}. 
The corresponding Einstein's equations of motion, resulting from variations with respect to $g_{\m\n}$ and $f_{\m\n}$ are
\begin{subequations}
\begin{align}
R_{ \mu\nu}(g)-\frac{1}{2}g_{ \mu\nu}R(g) & = {\frac{1}{2m_g^2}}T_{ \mu\nu} \,, \\
R_{ \mu\nu}(f)-\frac{1}{2}f_{ \mu\nu}R(f) & ={\frac{1}{2m_f^2}}\overline{T}_{ \mu\nu}\,,
\end{align}\label{EINab}
\end{subequations}
where $T_{ \mu\nu}$ and $\overline{T}_{ \mu\nu}$ are the two energy-momentum tensors
\begin{subequations}
\begin{align}
\label{TENSORSa}
T_{ \mu\nu}= & \,\partial_{ \mu}\phi\partial_{ \nu}\phi-\frac{1}{2}g_{ \mu\nu}(\partial\phi)^2\,-g_{ \mu\nu}\frac{m_g^2}{2\gamma}\left(1-e^{-\frac{\phi}{m_g\sqrt{3}}}\right)^2  \nonumber
\\ & +2m_g^2\,m^2
 \sum_{n=0}^{3} \beta_n e^{\frac{(n-4)\phi}{2\sqrt{3}m_g}}e^{-\frac{n\overline{\phi}}{2m_f\sqrt{3}}}V_{ \mu\nu}^{(n)}\,, \\[0.5cm]
 \overline{T}_{ \mu\nu}= & \,\partial_{ \mu}\overline{\phi}\partial_{ \nu}\overline{\phi}-\frac{1}{2}f_{ \mu\nu}(\partial\overline{\phi})^2\,-f_{ \mu\nu}\frac{m_f^2}{2\delta}\left(1-e^{-\frac{\overline{\phi}}{m_f\sqrt{3}}}\right)^2 \nonumber
 \\ & +2m_g^2\,m^2\sum_{n=1}^{4}\beta_n e^{\frac{(n-4)\phi}{2\sqrt{3}m_g}}e^{-\frac{n\overline{\phi}}{2m_f\sqrt{3}}} \tilde{V}_{ \mu\nu}^{(n)}\,.
 \label{TENSORSb} 
\end{align}
\end{subequations}
The tensors $V_{ \mu\nu}^{(n)}$ and $\tilde{V}_{ \mu\nu}^{(n)}$ are given in the Appendix~\ref{appendix_1}.  Varying with respect to $\phi$ and $\overline{\phi}$
we obtain the scalar field equations of motion which are given by
\begin{subequations}
\begin{align}
&\Box_g\phi+\frac{\partial}{\partial\phi}\left\{\frac{m_g^2}{2\gamma}\left(1-e^{-\frac{\phi}{m_g\sqrt{3}}}\right)^2\right\}
\\ &-  2m^2m_g^2\sum_{n=0}^{4} \b_n \frac{(n-4)}{2\sqrt{3}m_g}e^{\frac{(n-4)\phi}{2\sqrt{3}m_g}}e^{-\frac{n\overline{\phi}}{2m_f\sqrt{3}}}e_n(\sqrt{\Delta})=0\,,    \nonumber
\\[0.5cm] & \Box_f\overline{\phi}+\frac{\partial}{\partial\overline{\phi}}\left\{\frac{m_f^2}{2\delta}\left(1-e^{-\frac{\overline{\phi}}{m_f\sqrt{3}}}\right)^2\right\}
\\ & +  m^2\frac{m_g^2}{ m_f\sqrt{3}}\sum_{n=0}^4\beta_n
ne^{\frac{(n-4)\phi}{2\sqrt{3}m_g}}e^{-\frac{n\overline{\phi}}{2m_f\sqrt{3}}}e_n(\sqrt{\Delta})=0\,, \nonumber
\end{align} {\label{SCALAR EQS}}
\end{subequations}
 where the $\Box_{g,f}$ indicate the d'Alembert operator constructed from covariant derivatives of the corresponding metric tensor.

\section{Linearization}
\label{linearization}
The linear limit of the theory corresponds to considering the metrics to be perturbations around a common background metric, namely
\be g_{ \mu\nu}\,\approx\,\tilde{g}_{ \mu\nu}\,+\,h_{ \mu\nu},\,\,\,\,\,f_{ \mu\nu}\,\approx\,\tilde{g}_{ \mu\nu}\,+\,l_{ \mu\nu}\,.\ee
To zeroth order, the Einstein equations~\eqref{EINab} reduce to
\begin{subequations}
\begin{align}
R_{ \mu\nu}(\tilde{g})-\frac{1}{2}\tilde{g}_{ \mu\nu}R(\tilde{g}) & = {\frac{1}{2m_g^2}}T_{ \mu\nu}\,,  \\
R_{ \mu\nu}(\tilde{g})-\frac{1}{2}\tilde{g}_{ \mu\nu}R(\tilde{g}) & ={\frac{1}{2m_f^2}}\overline{T}_{ \mu\nu}\,,
\label{einstein}
\end{align}
\end{subequations}
dictating that the right-hand sides have to be equal and the energy-momentum tensors proportional. It is not difficult to see that in the case of equal quadratic curvature couplings, namely $\gamma=\delta$,
there is the following solution of the scalar equations of motion~\eqref{SCALAR EQS}:
\be \overline{\phi}\,=\,\frac{m_f}{m_g}\phi\,,{\label{SOL}}\ee 
provided that the constraint 
\be \sum_{n=0}^{4}\beta_n\left(n(  1+\alpha^{-2})-4\right)\varepsilon_n=0\,,{\label{CONSTR-1}}\ee  
is satisfied ($\alpha\equiv m_f/m_g$). The constants $\varepsilon_n$ are just $\varepsilon_n=e_n(1)=(1,4,6,4,1)$. In this case the energy-momentum tensors become proportional as $ T_{ \mu\nu} =\overline{T}_{ \mu\nu}/\a^2 $. Note that the constraint~\eqref{CONSTR-1} is immediately satisfied by a generalization of the so-called {\textit{minimal choice}} of parameters~\cite{Hassan:2011vm}, namely $\beta_n=\tilde{\beta}\left(3,\,-1,\,0,\,0,\,1\right)$, independent of $\alpha$, since both $\sum_{n=0}^{4}\varepsilon_n\beta_n$ and $\sum_{n=0}^{4}\varepsilon_n\beta_n n$ vanish.

Note that we could have expressed the equations in terms of the fields
\be
\Phi=\frac{m_g\phi+m_f\overline{\phi}}{\sqrt{m_g^2+m_g^2}}\quad \text{and}\quad\overline{\Phi}=\frac{m_f\phi-m_g\overline{\phi}}{\sqrt{m_g^2+m_f^2}}\,.{\label{BIGFS}}\ee
Then, the solution $\overline{\phi}=\alpha\phi$, existing for a common background, corresponds to $\overline{\Phi}=0,\, \forall \Phi$, provided the above constraint holds.
In the next section we show that the scalar system evolves rapidly toward $\overline{\Phi}=0$ and the system becomes essentially a system of one scalar mode that can be identified as a slow-rolling inflaton.

Therefore, assuming $\g=\d$ and the solution~\eqref{SOL}, for a common background, the energy-momentum tensors~\eqref{TENSORSa} and~\eqref{TENSORSb} become proportional, their proportionality resting on the relation
\be \sum_{ n=1}^{4}\beta_n \tilde{V}_{ \mu\nu}^{(n)}\,=\,\alpha^2\sum_{n=0}^{ 3}\beta_n V_{\mu\nu}^{(n)}\,,\ee
which is not a new constraint, since it reproduces the constraint ({\ref{CONSTR-1}}). This follows from 
\be V_{ \mu\nu}^{(n)}=\tilde{g}_{ \mu\nu}\pi_n,\,\,\,\,\,\,\tilde{V}_{ \mu\nu}^{(n)}=\tilde{g}_{ \mu\nu}\pi_{4-n}\,,\ee
with $\pi_n=(1,3,3,1,0)$. Then, in what follows we shall restrict our analysis to equal couplings (and, therefore, proportional energy-momentum tensors at the common background), keeping the constraint~\eqref{CONSTR-1}, which can be thought as fixing one of the $\beta$ parameters.

The full linearized action up to second order is
\begin{align}
\mathcal{S} &  = \int  {\rm d}^4x  \sqrt{-\tilde{g}} \bigg\{ \frac{1}{4} m_g^2h_{ \mu\nu}\tns{\mathcal{E}}{^\m^\n_\r_\s}h^{ \rho\sigma}+\frac{1}{4}m_f^2l_{ \mu\nu}\tns{\mathcal{E}}{^\m^\n_\r_\s}l^{ \rho\sigma} \nonumber
\\ & +  B(\Phi) \left( (h_{ \mu\nu}-l_{ \mu\nu})^2-(h-l)^2 \right) \nonumber
\\ & +A_g(\Phi) \left(h_{ \mu\nu}h^{ \mu\nu}-\frac{1}{2}h^2-2h-4\right) \nonumber
\\ & +A_f(\Phi)\left(l_{ \mu\nu}l^{ \mu\nu}-\frac{1}{2}l^2-2l-4\right) \nonumber
\\ & -\frac{1}{2(1+\a)^2}\partial_\m\Phi \partial_\n\Phi \left(h^{\m\n}+\frac{1}{2}h h^{\m\n} +\tns{h}{^\m_\r} h^{\n\r} \right) \nonumber
\\ & -\frac{\a^2}{2(1+\a)^2}\partial_\m\Phi \partial_\n\Phi \left(l^{\m\n}+\frac{1}{2}l l^{\m\n} +\tns{l}{^\m_\r} l^{\n\r} \right) \bigg\} \,.
\end{align}
Note that the Fierz-Pauli structure of the spin-$2$ mass term is automatic.  ${\cal{E}}_{ \mu\nu}^{\,\,\,\,\rho\sigma}$ is the Lichnerowicz operator and the appearing functions $A_{g,f}(\Phi)$ and $B(\Phi)$ are\footnote{ Note that $\sum_{n=0}^{4}\beta_n\pi_n=\alpha^{-2}\sum_{n=0}^{4}\beta_n\pi_{4-n}$ is identical to the constraint~\eqref{CONSTR-1}. }
\begin{subequations}
\begin{align}
A_{g}(\Phi)= \, &\frac{1}{8(1+\a)^2}(\tilde{\nabla}\Phi)^2+\frac{m_g^2}{8\gamma}\left(1-e^{-\frac{\Phi}{m_g\sqrt{3(1+\a)^2}}}\right)^2 \nonumber
\\  -&\frac{1}{2}m^2m_g^2e^{-\frac{2\Phi}{m_g\sqrt{3(1+\a)^2}}}\sum_n\beta_n{\pi_n}\,,
\\[0.5cm] A_{f}(\Phi)= \, & \a^2 \left(\frac{1}{8 (1+\a)^2}(\tilde{\nabla}\Phi)^2+\frac{m_g^2}{8\gamma}\left(1-e^{-\frac{\Phi}{m_g\sqrt{3(1+\a)^2}}}\right)^2\right. \nonumber
\\  -& \left.\frac{1}{2\alpha^2}m^2m_g^2e^{-\frac{2\Phi}{m_g\sqrt{3(1+\a)^2}}}\sum_n\beta_n{\pi_{4-n}} \right)\,,
\\[0.5cm]  B(\Phi) = \, &\frac{1}{4}m_g^2m^2 e^{-\frac{2\Phi}{m_g\sqrt{3(1+\a)^2}}}\left(\b_1+2\b_2+\b_3\right)\,.
\end{align}
\end{subequations}
The equality $A_f(\Phi)=\alpha^2A_g(\Phi)$ is valid upon imposing the constraint ({\ref{CONSTR-1}}).

We can then rewrite the action in terms of the mass eigenstates
\begin{subequations}
\begin{align}
M_{ \mu\nu}=\, &\frac{m_gm_f}{\sqrt{m_f^2+m_g^2}}\left(l_{ \mu\nu}-h_{ \mu\nu}\right)\,,\\
G_{ \mu\nu}=\, &\frac{1}{\sqrt{m_g^2+m_f^2}}\left(m_g^2h_{ \mu\nu}+m_f^2l_{ \mu\nu}\right)\,.    
\end{align}
\end{subequations}
Before we proceed to do so, notice that, if we assume that matter couples exclusively to one of the metrics\footnote{The issue of which metric tensor should couple to the Standard Model particle spectrum has been discussed extensively. It turns out that only two options are free of ghost degrees of freedom, one being the exclusive coupling to one of the metrics, namely $g_{ \mu\nu}$. The other option consists on coupling matter to a particular effective metric formed by the combination of the two metrics~\cite{Gumrukcuoglu:2015nua} (see also~\cite{deRham:2014naa, deRham:2014fha, Melville:2015dba}).}, namely, $g_{ \mu\nu}$, we have
\be g^{ \mu\nu}T_{ \mu\nu}^{(m)}\,\sim\,\frac{1}{\sqrt{m_g^2+m_f^2}}G^{ \mu\nu}T_{ \mu\nu}^{(m)}-\frac{\a}{\sqrt{m_g^2+m_f^2}}M^{ \mu\nu}T_{ \mu\nu}^{(m)}\,.\ee 
Then, we see that the massless graviton couples with the physical Planck mass, identified to be $M_{\rm P }^2/2\,=\,m_g^2+m_f^2$, while the massive spin-$2$ field couples with an extra factor of $\alpha\,=\,m_f/m_g$.

The linearized action written in terms of mass-eigenstates, and for $\overline{\Phi}=0$, is
\begin{align}
&\mathcal{S}   =  \int {\rm d}^4x \sqrt{-\tilde{g}} \Bigg\{ \frac14 G_{\m\n} \tns{\mathcal{E}}{^\m^\n_\r_\s}G^{\r\s}  + \frac14 M_{\m\n} \tns{\mathcal{E}}{^\m^\n_\r_\s}M^{\r\s} \nonumber
\\ & -\frac{1}{4}m_{\rm FP}^2(\Phi)  (M_{\m\n}M^{\m\n}-M^2) -4(1+\a^2)A_g(\Phi)\times \nonumber
\\ &  \bigg[1+ \frac{\sqrt{2}}{2M_{\rm P }}G-\frac{1}{2M_{\rm P }^2} \bigg(G_{\m\n}G^{\m\n} +M_{\m\n}M^{\m\n} -\frac{G^2}{2} -\frac{M^2}{2} \bigg) \bigg]\nonumber
\\ &   -\frac{\sqrt{2}}{2M_{\rm P }} \partial_\m\Phi \partial_\n\Phi \bigg[ G^{\m\n} +\frac{\sqrt{2}}{2M_{\rm P }}G^{\m\n}G +\frac{\sqrt{2}}{M_{\rm P }}\tns{G}{^\m_\r}G^{\n\r}\nonumber
\\ &   +\frac{\sqrt{2}}{2M_{\rm P }}M^{\m\n}M +\frac{\sqrt{2}}{M_{\rm P }}\tns{M}{^\m_\r}M^{\n\r}\bigg] \Bigg\}\,.
\end{align}
Note that the scalar $\Phi$ couples with $\mathcal{O}(1/M_{\rm P })$ couplings only to $G_{ \mu\nu}$, while all its $\mathcal{O}(1/M_{\rm P }^2)$ couplings are symmetric in $G_{ \mu\nu}$ and $M_{ \mu\nu}$.

The Fierz-Pauli mass of the massive spin-$2$ field is
\be m_{\rm FP}^2(\Phi)=-\frac{1}{2}M_{\rm P }^2 e^{-2\sqrt{\frac{2}{3}}\frac{\Phi}{M_{\rm P}}}\left(\b_1+2\b_2+\b_3\right)\,.
\ee
Note that we have set the redundant scale $m=m_f$. Note also that the spin-$2$ mass does not depend on the parameters $\beta_0,\,\beta_4$, being free of the constraint~\eqref{CONSTR-1}.

\begin{figure}[t!]
\centering
\includegraphics[width=0.4\textwidth]{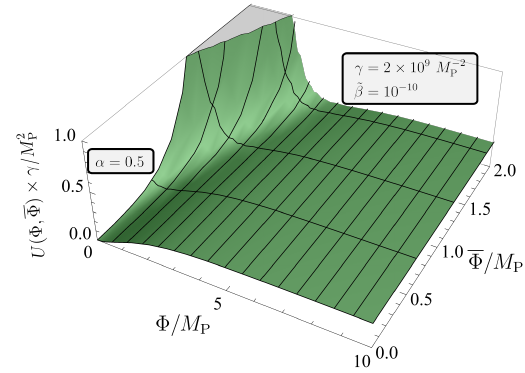}
\caption{The two-field potential given by~\eqref{eq:2field_pot_2} for $\a=0.5$ and $\tilde{\b}=10^{-10}$. The parameter $\gamma$ is chosen in such a way that the model remains consistent with the observations, as elaborated in Sec.~\ref{inflation}.} 
\label{fig:3D_pot}
\end{figure}
\section{The scalar system}
\label{the_scalar_system}
 Before delving into the analysis of inflation, let us take a step back and examine the scalar system without introducing the solution $\overline{\Phi}=0$.
The scalar potential at zeroth order, expressed in terms of the fields $\Phi$ and $\overline{\Phi}$, reads
\begin{align}
\label{eq:2field_pot_2}
 U(\Phi,\overline{\Phi}) &= \frac{M_{\rm P}^2}{4\g(1+\a^2)} \Bigg[\left(1-e^{-\sqrt{\frac{2}{3}} \frac{\Phi+\a \overline{\Phi}}{M_{\rm P}}} \right)^2 \nonumber 
\\& +  \a^2 \left(1-e^{-\sqrt{\frac{2}{3}} \frac{\Phi-\a^{-1} \overline{\Phi}}{M_{\rm P}}} \right)^2  \Bigg] 
\\ &  - \frac{\a^2 M_{\rm P}^4}{2(1+\a^2)^2} e^{-2\sqrt{\frac23} \frac{\Phi}{M_{\rm P}}} \sum_{n=0}^{4}\beta_n\varepsilon_ne^{\frac{n+\a^2(n-4)}{4\a} 2\sqrt{\frac23}\frac{\overline{\Phi}}{M_{\rm P}}}\,, \nonumber
\end{align}
where we have used also the relation $\pi_n+\pi_{4-n}=\varepsilon_n$. Differentiating with respect to $\overline{\Phi}$ we obtain
\begin{equation}
    \left.\frac{\partial U}{\partial\overline{\Phi}}\right|_{\overline{\Phi}=0}\,=\,0\,,
\end{equation}
thanks to the constraint~\eqref{CONSTR-1} that annihilates the contribution of the last term. The $\overline{\Phi}$-mass at the origin is
\begin{equation}
    \left.\frac{\partial^2U}{\partial\overline{\Phi}^2}\right|_{\Phi=\overline{\Phi}=0}=\frac{1}{3\gamma}\,-\frac{M_{\rm P}^2}{12(1+\alpha^2)^2}\sum_{n=0}^{4}\beta_n\varepsilon_n\left(n+\alpha^2(n-4)\right)^2\,,\end{equation}
    which has to be positive in order to avoid a tachyonic instability and, therefore, introduces a further constraint. If we employ the generalized {\textit{minimal choice}} $\beta_n=\tilde{\beta}(3,\,-1,\,0,\,0,\,1)$, reduces to
    \begin{equation}
     M^2_{\overline{\Phi}}\,=\,\frac{1}{3\gamma}-\tilde{\beta}M_{\rm P}^2\,\Longrightarrow\,\,\gamma^{-1}\geq 3\tilde{\beta}M_{\rm P}^2\,.{\label{CONSTR-2}}
    \end{equation}

To get some idea for the behavior of the potential in terms of $\overline{\Phi}$ we consider the slice $\Phi=0$ and obtain the leading large $\overline{\Phi}$ behavior to be
 \begin{equation}
    U(\Phi=0,\overline{\Phi})\approx\frac{\alpha^2M_{\rm P}^2}{4(1+\alpha^2)}\left(\frac{1}{\gamma}-\frac{2\beta_4 M_{\rm P}^2}{(1+\alpha^2)}\right)e^{\frac{2}{\alpha}\sqrt{\frac{2}{3}}\frac{\overline{\Phi}}{M_{\rm P}}}\,,
\end{equation}
which corresponds to an exponential growth and is positive in agreement with the constraint~\eqref{CONSTR-2}.

\begin{figure}[t!]
\centering
\includegraphics[width=0.4\textwidth]{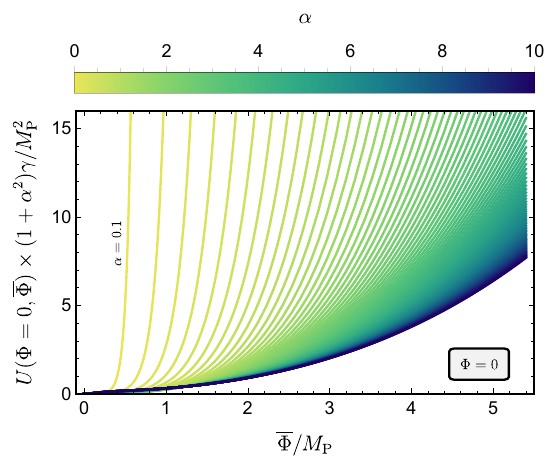}
\caption{The potential~\eqref{eq:2field_pot_2} in the $\Phi=0$ direction for various values of the parameter $\alpha$. The parameters $\tilde{\b}$ and $\g$ coincide with the ones shown in Fig.~\ref{fig:3D_pot}. } 
\label{fig:not_inf_pot}
\end{figure}

In Fig.~\ref{fig:3D_pot} is illustrated the two-field potential given by Eq.~\eqref{eq:2field_pot_2} for $\a=0.5$. It is clear that $\overline{\Phi}$ has a big slope toward $\overline{\Phi}=0$. The big slope along the $\overline{\Phi}$ direction arises due to the term $e^{+\frac{\overline{\Phi}}{\alpha M_{\rm P}}}$, which becomes even more important for $\a \ll 1$. This can also be seen from Fig.~\ref{fig:not_inf_pot}, where the slice $\Phi=0$ is depicted for different values of the parameter $\a$. As the value of $\a$ increases, the significance of the slope diminishes.  

In the upcoming section, we will see that the parameter $\g$ is linked to the CMB observations, compelling its value to be $\g = 2\times 10^{9}\, M_{\rm P}^{-2}$. Hence, we have opted for the overall factor $\tilde{\b} = 10^{-10}$ to ensure consistency with the constraint~\eqref{CONSTR-2}.

\section{The Inflaton Potential}
\label{infl_potential}
Having reduced the system to one-scalar field system we have that at zeroth order (i.e. for $G,\,M\rightarrow 0$) the scalar Lagrangian is
\be
\mathcal{L}(\Phi,\partial\Phi)= \sqrt{-\tilde{g}} \left(-\frac{1}{2}(\tilde{\nabla}\Phi)^2-U(\Phi) \right)\,,
\ee
with

\begin{align}
\label{eq:inf_pot}
U(\Phi)= & \,\frac{M_{\rm P}^2}{4\gamma}\left(1-e^{-\sqrt{\frac{2}{3}}\frac{\Phi}{M_{\rm P}}}\right)^2  \nonumber
\\& -\frac{M_{\rm P}^4}{2} \frac{\a^2}{1+\a^2}e^{-2\sqrt{\frac{2}{3}}\frac{\Phi}{M_{\rm P}}}\sum_{n=0}^{4}\beta_n \pi_n\,,
\end{align}
where we have used the constraint~\eqref{CONSTR-1} to substitute $\sum_{n=0}^{4}\beta_n \varepsilon_n \rightarrow (1+\a^2) \sum_{n=0}^{4}\beta_n \pi_n $.
The potential $U(\Phi)$ consists of a term having the standard Starobinsky functional form plus a term which, due to its exponential suppressing factor, is not expected to modify the inflationary behavior but could, in general, affect the behavior at the origin. The resulting minimum lies at
\be \Phi_0=M_{\rm P}\sqrt{\frac32}\ln\left[1-2\gamma M_{\rm P}^2 \frac{\a^2}{1+\a^2}\sum_{n=0}^{4}\beta_n\pi_n\right]\,,\ee 
with the potential value at the minimum being 
\be U(\Phi_0)=\frac{-\frac{M_{\rm P}^4}{2} \frac{\a^2}{1+\a^2} \sum_{n=0}^{4}\beta_n \pi_n}{1-2\g M_{\rm P}^2\frac{\a^2}{1+\a^2}\sum_{n=0}^{4}\beta_n \pi_n}\,.
\ee
The field-dependent scalar mass, defined as $m^2_{\Phi}(\Phi)=\frac{\partial^2 U}{\partial\Phi^2}$ is
\be
m_\Phi^2(\Phi) = \frac{e^{-2\sqrt{\frac{2}{3}}\frac{\Phi}{M_{\rm P}}}}{3\g} \left(2-e^{\sqrt{\frac{2}{3}}\frac{\Phi}{M_{\rm P}}} - \frac{4\g M_{\rm P}^2\a^2}{1+\a^2} \sum_{n=0}^{4}\beta_n \pi_n \right)  \,.
\ee

\begin{figure}[t!]
\centering
\includegraphics[width=0.4\textwidth]{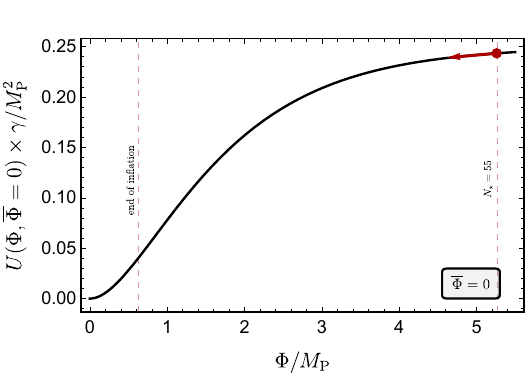}
\caption{The inflationary potential~\eqref{eq:inf_pot} with the characteristic plateau given by~\eqref{eq:2field_pot_2} for $\overline{\Phi}=0$. Note also that we have assumed $\sum_{n=0}^{4}\beta_n \pi_n=0$. } 
\label{fig:inf_pot}
\end{figure}

Adopting again the generalized {\textit{minimal choice}} of the parameters $\beta_n=\tilde{\beta}(3,\,-1,\,0,\,0,\,1)$ we see in addition to the constraint~\eqref{CONSTR-1} that they also yield
\begin{equation}
 \sum_{n=0}^{4}\beta_n\pi_n=0\,.   
\end{equation}
Therefore, this choice of parameters kills the second term in the potential rendering a potential to 
\be U(\Phi)\,=\,\frac{M_{\rm P}^2}{4\gamma}\left(1-e^{-\sqrt{\frac{2}{3}}\frac{\Phi}{M_{\rm P}}}\right)^2\,,\ee
which is identical to the standard Starobinsky model~\cite{Starobinsky1980}\footnote{ A key feature of the Starobinsky potential, critical for the accomplishment of slow-roll inflation, is its asymptotic behavior, i.e. $U(\Phi) \rightarrow \text{const.} $ as $\Phi \rightarrow \infty $. In broader models like $F(R) = a_1 R + a_2 R^2 + \ldots + a_n R^n$ theories, the trend is, $U(\Phi) \rightarrow 0 $ as $\Phi$ extends to infinity. Consequently, the absence of an infinite plateau in these models prevents inflation to occur across a wide range of initial conditions. } as can also be seen from Fig.~\ref{fig:inf_pot}. For this choice the minimum lies at the origin, i.e. $\Phi_0=0$ and $U(\Phi_0)=0$ in agreement with a vanishing cosmological constant. The scalar mass at the origin is 
\be m_{\Phi}^2(0)=\frac{1}{3\gamma}\,.\ee
The Fierz-Pauli mass at the minimum is
\be m_{\rm FP}^2(0)=-M_{\rm P }^2\left(\beta_1+2\beta_2+\beta_3\right)/2\,.\ee 
Note that it depends on the three parameters $\beta_1,\,\beta_2,\,\beta_3$, while $\beta_4$ is determined by the condition ({\ref{CONSTR-1}}) and $\beta_0$ by the condition on zero-cosmological constant $\beta_0=-3(\beta_1+\beta_2)-\beta_3$.
The generalized \textit{minimal choice} of parameters $\beta_n=\tilde{\beta}\left(3,\,-1,\,0,\,0,\,1\right)$ gives a Fierz-Pauli mass
\be m_{\rm FP}^2(0)=\frac{3}{2}\tilde{\beta}M_{\rm P }^2\,.\ee
Nevertheless, a more general choice of the $\beta_n$'s is allowed that could lead to a value quite lighter than $M_{\rm P }$ and phenomenologically more interesting. For instance, the overall scale $\tilde{\beta}$ could have been much lower than its upper bound $\mathcal{O}((\gamma M_{\rm P}^2)^{-1})$ leading to a much lighter $m_{\rm FP}$ independent of the bounds set for $\alpha$.

\section{Inflation}
\label{inflation}
Regarding the cosmological observables and under the assumption of the slow-roll approximation, we start by examining the scalar and tensor power spectra, which hold significant importance in the study of inflationary cosmology. By selecting an arbitrary pivot scale $k_\star$ that exited the horizon, we can express the scalar power spectrum ($\mathcal{P}_\zeta $) and the tensor power spectrum ($\mathcal{P}_T$) as follows:
\begin{equation}
\mathcal{P}_\zeta (k)=A_s \left(\frac{k}{k_\star} \right)^{n_s -1},  \quad  \mathcal{P}_T (k)\simeq\frac{2U(\Phi_\star)}{3\pi^2} \left(\frac{k}{k_\star} \right)^{n_t}\,,
\end{equation}
where $\Phi_\star$ is the field value when the pivot scale, $k_\star =a H$, left the Hubble radius. In the slow-roll approximation the amplitude of the scalar power spectrum is given by
\be
\label{eq:As}
A_s\simeq\frac{1}{24\pi^2 \epsilon_U(\Phi_\star)}\frac{U(\Phi_\star)}{M_{\rm P}^4} \simeq 2.1\times 10^{-9}\,,
\ee
at the pivot scale $k_\star = 0.05 \, {\rm Mpc}^{-1}$ as indicated by the latest Planck 2018 data~\cite{Planck:2018jri}.
The scale dependence is given in terms of the scalar ($n_s$) and tensor ($n_t$) spectral indices, which are given by
\begin{equation}
n_s-1=\frac{{\rm d} \ln \mathcal{P}_\zeta (k) }{{\rm d} \ln k}\simeq -6\epsilon_U +2\eta_U\,,\quad n_t= \frac{{\rm d} \ln \mathcal{P}_T (k) }{{\rm d} \ln k}\,,
\end{equation}
where we have used the potential slow-roll parameters
\begin{equation}
\epsilon_U=\frac{M_{\rm P}^2}{2}\left(\frac{U'(\Phi)}{U(\Phi)} \right)^2 \quad \mbox{and} \quad \eta_U = M_{\rm P}^2\frac{U''(\Phi)}{U(\Phi)}\,.
\end{equation}
The tensor-to-scalar ratio is defined as
\begin{equation}
    r= \frac{\mathcal{P}_T (k)}{\mathcal{P}_\zeta (k)} \simeq 16\epsilon_U.
\end{equation}
The recent release of the BICEP/Keck~\cite{BICEP:2021xfz} data, imposes the bound $r<0.036$
at the 95\% C.L. at the pivot scale $0.05 \, {\rm Mpc}^{-1}$, while the scalar spectral index is restricted to be $n_s = 0.9649\pm0.0042$~\cite{Planck:2018jri}.

The model under consideration exhibits the same inflationary observables as the known Starobinsky model of inflation~\cite{Starobinsky1980} with the characteristic plateau-type potential at large field values. 

To quantify the allowed range of the parameters we need an estimate for $\Phi_\star$. To this
goal we use the number of $e$-folds ($N_\star$), left to the end of inflation, which is given by
\be
\label{eq:Nstar}
N_\star = \int_{\Phi_{\rm end}}^{\Phi_\star} \frac{{\rm d}\Phi}{M_{\rm P}\sqrt{2\epsilon_U}}\simeq \frac34 e^{\sqrt{\frac23} \frac{\Phi_\star}{M_{\rm P} }}\,,
\ee
given that $\Phi_\star \gg \Phi_{\rm end}$. Solving~\eqref{eq:Nstar} with respect to the field value at the pivot scale we obtain that $\Phi_\star \stackrel{\scriptscriptstyle N_\star=55}{\simeq} 5.26 \,M_{\rm P}$. The field value at the end of inflation is given by the condition $\epsilon_H = -\dot{H}/H^2=1$. At first order in the potential slow-roll parameters, this is achieved when $\epsilon_U \simeq \left(1+\sqrt{1-\eta_U/2}\right)^2$, from which we obtain that $\Phi_{\rm end} \simeq 0.63\, M_{\rm P} $. Using $\Phi_\star$, given above, the amplitude of the scalar power spectrum~\eqref{eq:As} is written, in terms of the number of $e$-folds, as
\be
A_s = \frac{\sinh^4[\Phi_\star/(\sqrt{6}M_{\rm P})]}{8\pi^2 M_{\rm P}^2 \gamma} \simeq \frac{N_\star^2}{72\pi^2M_{\rm P}^2\gamma}\,,
\ee
from which we obtain that
$\gamma \stackrel{\scriptscriptstyle N_\star=55}{\simeq} 2\times 10^9\, M_{\rm P}^{-2}$.  The value of the parameter $\gamma$ aligns with that of the original Starobinsky model~\cite{Starobinsky1980}.
Finally, the scalar spectral index and the tensor-to-scalar ratio are calculated to be
\be 
n_s = 1 -\frac{2}{N_\star} \stackrel{\scriptscriptstyle N_\star=55}{\simeq} 0.9636\,, \quad r= \frac{12} {N_\star^2} \stackrel{\scriptscriptstyle N_\star=55}{\simeq} 0.0040\,,
\ee
in agreement with the CMB observations~\cite{Planck:2018jri, BICEP:2021xfz}.

\section{Conclusions}
\label{conclusions}

In this article, we reconsidered the framework of the standard bimetric theory of gravity originally proposed by Hassan and Rosen~\cite{Hassan:2011zd}. This theory involves two distinct dynamical metrics resulting in describing apart from the standard massless graviton an additional massive spin-$2$ field in a fashion devoid of ghosts. We considered an extension of the corresponding action to include quadratic Ricci scalar terms, namely $\sqrt{-g}R^2(g)$ and $\sqrt{-f}R^2(f)$, associated with both metrics, in accordance with general arguments that such terms are expected to be generated by the quantum interactions of gravitating matter.

Our analysis at the linear level showed that the theory exhibits a particle spectrum that encompasses the familiar massless graviton, a massive spin-2 field with a mass of the Fierz-Pauli type as well as a pair of scalar fields, while all standard consistency requirements of the bimetric framework were kept, allowing, in general for a phenomenologically interesting interpretation of the additional fields. In particular, the scalar sector of the model was shown to generate an inflaton potential that precisely mirrors the one found in the standard Starobinsky inflationary model~\cite{Starobinsky1980}. Therefore, the extended bimetric theory, with quadratic Ricci scalar terms, could, in principle, describe both dark matter and inflation.

\vspace{1.5 cm}
\paragraph*{ Acknowledgments.} 
IDG would like to dedicate this paper to the memory of his father, Dimitrios I. Gialamas, who fought hard throughout his life with the sole purpose of helping him to achieve his own goals.
The work of IDG was supported by the Estonian Research Council grant SJD18.

\appendix*
\section{Formulas}\label{appendix_1}

The tensors $V_{ \mu\nu}^{(n)}$ and $\tilde{V}_{ \mu\nu}^{(n)}$  appearing in the energy-momentum tensors~\eqref{TENSORSa} and~\eqref{TENSORSb} are given by
\newpage
\begin{align}
&V^{(0)}_{\mu\nu} = g_{\mu\nu}\,, \nonumber
\\ &V^{(1)}_{\mu\nu} = g_{\mu\nu} {\rm Tr}\left[X\right] - g_{\nu\rho} \tns{X}{^\rho_\mu}\,, \nonumber
\\ &V^{(2)}_{\m\n} = g_{\n\r} \left( \tns{{X^2}}{^\rho_\mu} -  {\rm Tr}\left[X\right]  \tns{X}{^\rho_\mu} \right) +\frac{g_{\m\n}}{2} \left( {\rm Tr}\left[X\right]^2 -{\rm Tr}\left[X^2\right]\right)\,, \nonumber
\\ &V^{(3)}_{\mu\nu} = -g_{\n\r}\left( \tns{{X^3}}{^\r_\m} - {\rm Tr}\left[X\right] \tns{{X^2}}{^\r_\m} +\frac{1}{2} \left(  {\rm Tr}\left[X\right]^2 -{\rm Tr}\left[X^2\right]\right)   \tns{X}{^\r_\m} \right)\nonumber
\\& \,\,\,\,+ \frac{g_{\mu\nu}}{6} \left( {\rm Tr}\left[X\right]^3 -3 {\rm Tr}\left[X\right] {\rm Tr}\left[X^2\right] + 2{\rm Tr}\left[X^3\right] \right)\,,
\end{align}
and
\begin{align}
\tilde{V}^{(1)}_{\mu\nu} &= f_{\n\r} \tns{X}{^\rho_\mu}\,, \nonumber
\\ \tilde{V}^{(2)}_{\mu\nu} &= f_{\n\r} \left({\rm Tr}\left[X\right]  \tns{X}{^\r_\m} - \tns{{X^2}}{^\r_\m}\right) \,, \nonumber
\\ \tilde{V}^{(3)}_{\mu\nu} &= f_{\n\r}\left( \tns{{X^3}}{^\r_\m}  - {\rm Tr}\left[X\right]  \tns{{X^2}}{^\r_\m}  +\frac{1}{2} \left( {\rm Tr}\left[X\right]^2 - {\rm Tr}\left[X^2\right]\right)  \tns{X}{^\rho_\mu}\right)\,, \nonumber
\\ \tilde{V}^{(4)}_{\mu\nu} &= f_{\mu\nu}  \,,
\end{align}
where $X=\sqrt{\Delta}$.

\bibliography{bigravity_refs}

\end{document}